\documentclass[final,5p,times,twocolumn]{elsarticle}

\usepackage{\jobname}

\newif\iflongversion \longversionfalse

\journal{Information Processing Letters}

\begin{document}

\begin{frontmatter}
  \title{The Complexity of Propositional Implication\tnoteref{DFG}}

  \author[Hannover]{Olaf Beyersdorff}
  \ead{beyersdorff@thi.uni-hannover.de}

  \author[Hannover]{Arne Meier}
  \ead{meier@thi.uni-hannover.de}

  \author[Hannover]{Michael Thomas}
  \ead{thomas@thi.uni-hannover.de}

  \author[Hannover]{Heribert Vollmer}
  \ead{vollmer@thi.uni-hannover.de}
 
  \address[Hannover]{%
    Institut f\"ur Theoretische~Informatik, Gottfried Wilhelm Leibniz Universit\"{a}t, \\%
    Appelstr.~4, 30167~Hannover, Germany%
  }

  \tnotetext[DFG]{Research supported in part by DFG grant VO 630/6-1.}
  
  \begin{abstract}
    The question whether a set of formulae $\Gamma$ implies a formula $\varphi$ 
    is fundamental.
    The present paper studies the complexity of the above implication problem for 
    propositional formulae that are built from a systematically restricted set of Boolean 
    connectives. We give a complete complexity-theoretic classification for all sets of Boolean 
    functions in the meaning of Post's lattice and show that the implication problem 
    is efficiently solvable only if the connectives are definable using the 
    constants $\{\false,\true\}$ and only one of $\{\land,\lor,\xor\}$. The problem 
    remains $\co\NP$-complete in all other cases. We also consider the restriction 
    of $\Gamma$ to singletons which makes the problem strictly easier
    in some cases.
  \end{abstract}

  
\end{frontmatter}

\section{Introduction}

$\SAT$, the satisfiability problem for propositional formulae, is the most fundamental and historically
the first $\NP$-complete problem (proven by S.~Cook and L.~Levin \cite{coo71,lev73}). A natural question, posed by
H.~Lewis in 1979, is what the sources of hardness in the Cook-Levin Theorem are. More precisely, Lewis systematically restricted the language of propositional formulae and determined the computational complexity of the
satisfiability problem depending on the set of allowed
connectives. E.\,g., if only logical ``and'' ($\wedge$) and ``or''
($\vee$) are allowed, we deal with \emph{monotone formulae} for which
the satisfiability problem obviously is easy to solve (in polynomial
time). Lewis proved that $\SAT$ is $\NP$-complete iff the negation of
implication, $x\wedge \neg y$, is among the allowed connectives or can be
simulated by the allowed connectives \cite{lew79}. To simulate a logical
connective $f$ by a set of logical connectives (or, in
other words, a set of Boolean functions) $B$ formally means that $f$
can be obtained from functions from $B$ by superposition, i.\,e.,
general composition of functions. Equivalently, we can express this fact
by saying that $f$ is a member
of the \emph{clone} generated by $B$, in symbols $f\in[B]$. 

This brings us into the realm of Post's lattice, the lattice of all Boolean
clones \cite{pos41}. In this framework, Lewis' result can be restated
as follows. Let $\SAT(B)$ denote the 
satisfiability problem for propositional formulae with connectives
restricted to the set $B$ of Boolean functions. Then $\SAT(B)$ is
$\NP$-complete iff $\CloneS_1\subseteq[B]$; otherwise the problem is
polynomial-time solvable. Note that the 2-ary Boolean function
$x\wedge\neg y$ forms a basis for $\CloneS_1$. 

Since then, many problems related to propositional formulae and
Boolean circuits have been studied for restricted sets of connectives
or gates, and their computational complexity has been classified,
depending on a parameter $B$, as just explained for $\SAT$. These
include, e.\,g., the equivalence problem \cite{rei03}, the circuit
value problem \cite{rewa05}, the quantified Boolean formulae problem
QBF \cite{rewa05}, but also more recent questions related to
non-classical logics like LTL \cite{bsssv07}, CTL \cite{memuthvo08},
or default logic \cite{BMTV09}. An important part of the proof of
the classification of different reasoning tasks for default logic in
the latter paper \cite{BMTV09} was the identification of the
$\co\NP$-complete and polynomial-time solvable fragments of the
\emph{propositional implication problem}. Though implication is
without doubt a very fundamental and natural problem, its
computational complexity has not yet been fully classified. This is
the purpose of the present note.  

We study the problem, given a set $\Gamma$ of propositional formulae
and a formula $\varphi$, to decide whether $\varphi$ is implied by
$\Gamma$. Depending on the set of allowed connectives in the occurring
formulae, we determine the computational complexity of this problem as
$\co\NP$-complete, $\ParityL$-complete, in
$\AC{0}\complexityClassFont{[2]}$, or in $\AC{0}$. The type of
reduction we use are \emph{constant-depth reductions} \cite{chstvi84}
and the weaker \emph{$\AC{0}$ many-one reductions}. For both
reductions, $\AC{0}$ forms the $\mathbf{0}$-degree. We also consider
the case of the problem restricted to singleton sets $\Gamma$, the
\emph{singleton-premise implication problem}. Interestingly, the complexity
of the previously $\ParityL$-complete cases now drops down to the
class $\AC{0}\complexityClassFont{[2]}$; in all other cases the
complexity remains the same as for the unrestricted problem. Finally, as an easy
consequence our results give a refinement of Reith's previous
classification of the equivalence problem for propositional formulae
\cite{rei03}. While Reith only considered the dichotomy between the
$\co\NP$-complete and logspace-solvable cases, we show that under
constant-depth reductions, three complexity degrees occur:
$\co\NP$-complete, membership in $\AC{0}\complexityClassFont{[2]}$, and
membership in $\AC{0}$.

\section{Preliminaries}

  In this paper we make use of standard notions of complexity
  theory. The arising complexity degrees encompass the classes 
  $\AC{0}$, $\AC{0}\complexityClassFont{[2]}$, $\ParityL$,
  $\P$, and $\co\NP$ (cf.~\cite{pap94,vol99} for background information).

  $\AC{0}$ forms the class of languages recognizable by logtime-uniform Boolean
  circuits of constant depth and polynomial size over
  $\{\land,\lor,\neg\}$, where the fan-in of gates of the first two types is not bounded.
  The class $\AC{0}\complexityClassFont{[2]}$ is defined similarly as
  $\AC{0}$, but in addition to $\{\land,\lor,\neg\}$ we also allow $\xor$-gates of unbounded fan-in.  
  The class $\ParityL$ is defined as the class of languages $L$ for
  which there exists a nondeterministic logspace Turing machine $M$
  such that for all $x$, $x \in L$ iff $M(x)$ has an odd number of accepting paths.

  For the hardness results we use \emph{constant-depth} and \emph{$\AC{0}$ many-one reductions},
  defined as follows:
  A language $A$ is \emph{constant-depth reducible} to a language
  $B$ ($A\leqcd B$) if there exists a logtime-unifordm $\AC{0}$-circuit family
  $\{C_n\}_{n\geq 0}$ with $\{\land,\lor,\neg\}$-gates and oracle gates
  for $B$ such that for all $x$, $C_{|x|}(x) = 1$ iff $x \in A$ \cite{vol99}.
  A language $A$ is \emph{$\AC{0}$ many-one reducible} to a language
  $B$ ($A\leqacm B$) if there exists a function $f$ computable by
  a logtime-uniform $\AC{0}$-circuit family
  such that $x \in A$ iff $f(x) \in B$.
  
  For both reductions, the class $\AC{0}$ forms the $\mathbf{0}$-degree. 
  Furthermore, it is easy to see that 
  $$
    \MOD_2 := \{ w \in \{\false,\true\}^\star
    \mid |w|_\true \equiv \true \pmod 2 \},  
  $$
  where $|w|_\true = |\{ i \mid 1 \leq i \leq n,\ w_i = \true \}|$, is complete for
  $\AC{0}\complexityClassFont{[2]}$ under $\leqcd$-reductions, for $\AC{0}\complexityClassFont{[2]}$ merely 
  extends $\AC{0}$ with oracle gates for $\MOD_2$.

  We assume familiarity with propositional logic. 
  The set of all propositional formulae is denoted by $\allFormulae$.
  For $\Gamma \subseteq \allFormulae$ and  $\varphi\in \allFormulae$, we
  write $\Gamma \models \varphi$ iff all assignments satisfying all
  formulae in $\Gamma$ also satisfy $\varphi$. 
  
\section{Boolean Clones} \label{sect:clones}

  In order to completely classify the complexity of the implication problem for all possible sets $B$
  of Boolean functions, one has to consider an infinite number of parameterized problems. 
  We introduce the notion of a clone to reduce the number of problems to be considered to a finite set.

\begin{figure*}[p]
    \centering
    \includegraphics[width=0.6\linewidth]{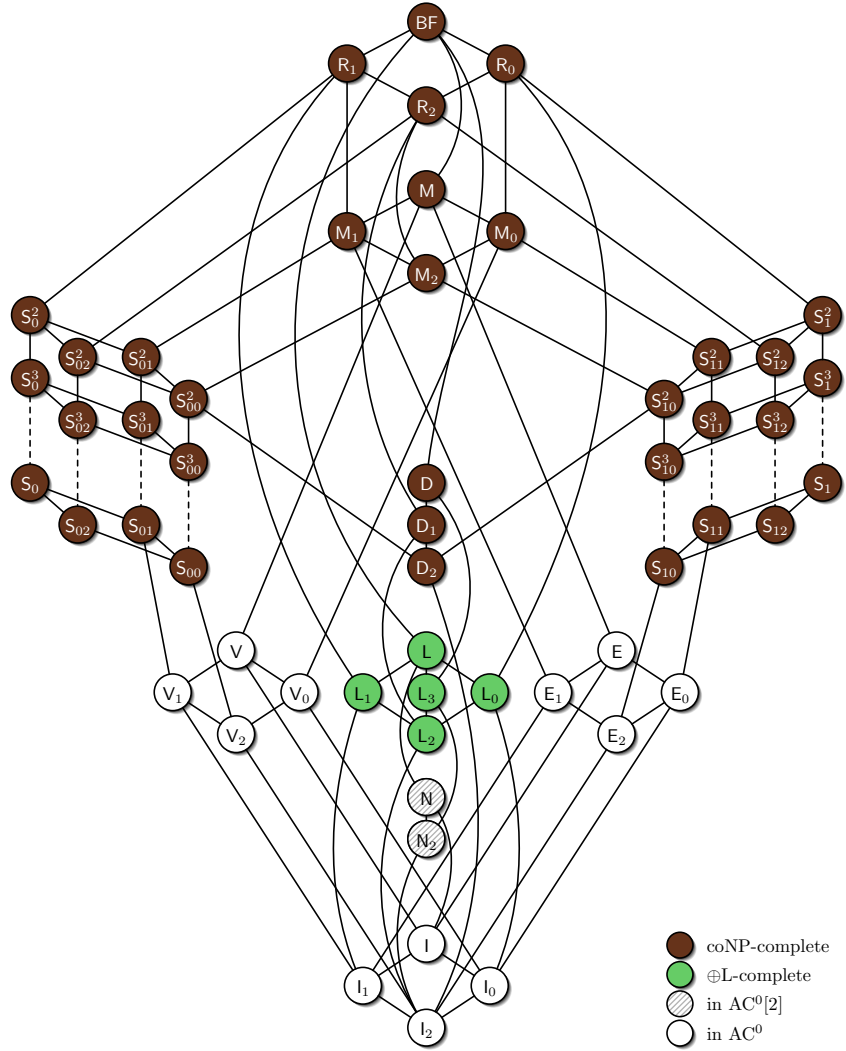}
    \caption{\label{fig:implication}%
    Post's lattice. Colors indicate the complexity of $\IMP(B)$, the implication problem for $B$-formulae.}
  \end{figure*}

  A propositional formula using only connectives from a finite set $B$
  of Boolean functions is called a $B$-formula. 
  The set of all $B$-formulae is denoted by $\allFormulae(B)$.
  A \emph{clone} is a set $B$ of Boolean functions that is closed under superposition, 
  i.\,e., $B$ contains all projections and is closed under arbitrary
  composition. 
  We denote by $[B]$ the smallest clone containing $B$ and call $B$ a \emph{base} for $[B]$.
  In \cite{pos41} Post classified the lattice of all clones and found a finite base for each clone, 
  see Fig. \ref{fig:implication}.
  In order to introduce the clones relevant to this paper, we define the following notions
  for $n$-ary Boolean functions $f$:  
  \begin{itemize} \itemsep 0pt 
    \item $f$ is \emph{$c$-reproducing} if $f(c, \ldots , c) = c$, $c \in \{\false,\true\}$.
    \item $f$ is \emph{monotone} if $a_1 \leq b_1, \ldots , a_n \leq b_n$ implies $f(a_1, \ldots , a_n) \leq f(b_1, \ldots , b_n)$.
    \item $f$ is \emph{$c$-separating} if there exists an index $i \in \{1, \ldots , n\}$ such that $f(a_1, \ldots , a_n) = c$ implies $a_i = c$, $c \in \{\false,\true\}$.
    \item $f$ is \emph{self-dual} if $f \equiv \dual{f}$, where $\dual{f}(x_1, \ldots , x_n) := \neg f(\neg x_1, \ldots , \neg x_n)$.
    \item $f$ is \emph{linear} if $f \equiv c_0 \xor c_1 x_1 \xor \cdots \xor c_n x_n \xor c$ for constants $c_i \in \{0, 1\}$, $0\leq i \leq n$, and variables $x_1, \ldots , x_n$.
  \end{itemize}
  The clones relevant to this paper are listed in Table \ref{tab:clones}. The definition of all Boolean clones can be found, e.\,g., in \cite{bcrv03}. 
    
  \begin{table*}
    \centering
    \begin{tabular}{c|l|l}
      Name & Definition & Base \\
      \hline
      $\CloneBF$ & All Boolean functions & $\{\land, \neg\}$ \\
      $\CloneM_2$ & $\{f : f \text{ is monotone and $0$- and $1$-reproducing}\}$ & $\{\lor, \land\}$ \\
      $\CloneS_{00}$ & $\{f : f \text{ is $0$-separating}\} \cap \CloneM_2$& $ \{x \lor (y \land z)\} $ \\
      $\CloneS_{10}$ & $\{f : f \text{ is $1$-separating}\} \cap \CloneM_2$ & $ \{x \land (y \lor z)\} $ \\
      $\CloneD_2$ & $\{f : f \text{ is monotone and self-dual}\}$ & $ \{(x\land y) \lor (y\land z) \lor (x\land z)\}$ \\
      $\CloneL$ & $\{f : f \text{ is linear}\}$ & $\{ \xor,\true\}$ \\
      $\CloneL_2$ & $\{f : f \text{ is linear and $0$- and $1$-reproducing}\}$ & $\{ x \xor y \xor z\}$ \\
      $\CloneV$ & 
      $\{f : f \equiv c_0 \lor \bigvee_{i=1}^n c_ix_i \text{ for } c_i \in \{\false,\true\}, 1 \leq i \leq n\}$
      & $\{ \lor, \false,\true \}$ \\
      $\CloneE$ & 
      $\{f : f \equiv c_0 \land \bigwedge_{i=1}^n c_ix_i \text{ for } c_i \in \{\false,\true\}, 1 \leq i \leq n\}$
      & $\{ \land, \false, \true \}$ \\
      $\CloneN$ & $\{f : f \text{ depends on at most one variable}\}$ & $\{ \neg,\true\}$ \\
      $\CloneN_2$ & $\{f : f \text{ is the negation or a projection}\}$ & $\{ \neg\}$ \\
    \end{tabular}
    \caption{
      \label{tab:clones}
      A list of Boolean clones with definitions and bases.
    }
  \end{table*}

\section{The Complexity of the Implication Problem}
  
  Let $B$ be a finite set of Boolean functions. 
  The \emph{implication problem} for $B$-formulae is defined as  
  
  \begin{center}
    \begin{tabular}{ll}
      \textit{Problem}:& $\IMP(B)$ \\
      \textit{Instance}:  & A finite set $\Gamma$ of $B$-formulae and a $B$-formula $\varphi$.\\
      \textit{Question}: & Does $\Gamma \models \varphi$ hold?
    \end{tabular}
  \end{center}
  
  In the general case $[B]=\CloneBF$, verifying an instance $(\Gamma,\varphi)\in\IMP(B)$ amounts to verifying that the formula $\bigwedge\!\Gamma \limplies \varphi$ is tautological. 
  We hence obtain a $\co\NP$ upper bound.
  The following theorem classifies the complexity of the implication problem for all possible sets $B$.
  
  \begin{theorem} \label{thm:implication}
    Let $B$ be a finite set of Boolean functions. Then the implication problem for propositional $B$-formulae, $\IMP(B)$, is
    \begin{enumerate}
      \item $\co\NP$-complete under $\leqacm$-reductions if $\CloneS_{00} \subseteq [B]$ or
        $\CloneS_{10} \subseteq [B]$ or $\CloneD_2 \subseteq [B]$, 
      \item $\ParityL$-complete under $\leqacm$-reductions if $\CloneL_2 \subseteq [B] \subseteq \CloneL$,
      \item in $\AC{0}\complexityClassFont{[2]}$ and $\MOD_2 \leqacm \IMP(B)$ if $\CloneN_2 \subseteq [B] \subseteq \CloneN$,
      and
      \item in $\AC{0}$ for all other cases.
    \end{enumerate}
  \end{theorem}
  
  In contrast to the first two cases, we do not state a
  completeness result for the third case, where $\CloneN_2 \subseteq
  [B] \subseteq \CloneN$. 
  Under $\leqcd$-reductions, however, $\IMP(B)$ is
  $\AC{0}\complexityClassFont{[2]}$-complete in this case.
  For $\leqacm$-reductions, the existence of a complete problem $A$
  would state that any $\AC{0}\complexityClassFont{[2]}$-circuit is
  equivalent to an $\AC{0}$-computation followed by a single oracle
  call to $A$. To date, there is no such problem known.

  We split the proof of Theorem~\ref{thm:implication} into several lemmas.
  
  \begin{lemma} \label{lem:implication-S00-and-S01}
    Let $B$ be a finite set of Boolean functions such that
    $\CloneS_{00} \subseteq [B]$ or $\CloneS_{10} \subseteq [B]$. 
    Then $\IMP(B)$ is $\co\NP$-complete under $\leqacm$-reductions.
  \end{lemma}

  \begin{proof}
    Membership in $\co\NP$ is apparent, because given $\Gamma$ and
    $\varphi$, we just have to check
    that for all assignments $\sigma$ to the variables of $\Gamma$ and
    $\varphi$, either $\sigma\not\models \Gamma$ or $\sigma\models
    \varphi$.
 
    The hardness proof is inspired by \cite{rei03}.
    Observe that $\IMP(B) \equivcd \IMP(B \cup \{\true\})$ if $\land
    \in [B]$, and that $\IMP(B) \equivcd \IMP(B \cup \{\false\})$ if
    $\lor \in [B]$ (because $\varphi \models \psi \iff \varphi_{[\true / t]}
    \land t \models \psi_{[\true / t]}$ and $\varphi \models
    \psi \iff \varphi_{[\false / f]} \models \psi_{[\false /
        f]} \lor f$ where $t,f$ are new variables). 
    It hence suffices to show that $\IMP(B)$ is $\co\NP$-hard for
    $\CloneM_0 = [\CloneS_{00} \cup \{\false\}]=[\{\land,\lor,\false\}]$ and $\CloneM_1 =
    [\CloneS_{10} \cup \{\true\}]=[\{\land,\lor,\true\}]$. We will show that $\IMP(B)$ is
    $\co\NP$-hard for each base $B$ with $\CloneM_2=[\{\land,\lor\}] \subseteq [B]$. 
    To prove this claim, we will provide a reduction from $\TAUTDNF$
    to $\IMP(B)$, where $\TAUTDNF$ is the $\co\NP$-complete problem
    to decide, whether a given propositional formula in disjunctive
    normal form is  a tautology. 
    
    Let $\varphi$ be a propositional formula in disjunctive normal
    form over the propositions $X=\{x_1,\ldots,x_k\}$. Then $\varphi =
    \bigvee_{i=1}^n\bigwedge_{j=1}^m l_{ij}$, where $l_{ij}$ are
    literals over $X$. 
    Take new variables $Y=\{y_1,\dots,y_k\}$ and replace in $\varphi$ each
    negative literal $l_{ij}=\neg x_l$ by $y_l$. 
    Define the resulting formula as $\psi_2$ and let $\psi_1 := \bigwedge_{i=1}^k(x_i \lor y_i)$.
    We claim that $\varphi \in \TAUTDNF \iff \psi_1 \models \psi_2$. 
    
    Let us first assume $\varphi \in \TAUTDNF$ and 
    let $\sigma\colon X  \cup Y \to \{\false,\true\}$ be an assignment
    such that $\sigma\models \psi_1$.
    As $\varphi$ is a tautology, $\sigma\models \varphi$.
    But also $\sigma \models \psi_2$, as we
    simply replaced the negated variables in $\varphi$ by positive
    ones and $\psi_2$ is monotone. It follows that $\psi_1 \models \psi_2$, 
    since $\sigma$ was arbitrarily chosen. 
    
    For the opposite direction, let $\varphi \notin \TAUTDNF$. Then
    there exists an assignment $\sigma\colon X \to \{\false, \true\}$
    such that $\sigma \not\models \varphi$. We extend $\sigma$ to an
    assignment $\sigma' \colon X
    \cup Y \to \{\false,\true\}$ by setting $\sigma'(y_i)=1-x_i$ for
    $i=1,\dots,k$. Then 
    $\sigma'(x_i)=\false$ iff $\sigma'(y_i)=\true$, and consequently
    $\sigma'$ simulates $\sigma$ on $\varphi'$. As a result, $\sigma'
    \not\models \psi_2$. Yet, either $\sigma'(x_i)=\true$ or
    $\sigma'(y_i)=\true$ for $i=1,\dots,k$. Thus $\sigma' \models \psi_1$, yielding
    $\psi_1 \not\models \psi_2$. 
  \end{proof}
  
  \begin{lemma} \label{lem:implication-D2}
    Let $B$ be a finite set of Boolean functions such that
    $\CloneD_{2} \subseteq [B]$. Then $\IMP(B)$ is 
    $\co\NP$-complete under $\leqacm$-reductions.
  \end{lemma}
  
  \begin{proof}
    Again we just have to argue for $\co\NP$-hardness of $\IMP(B)$.
    We give a reduction from $\TAUTDNF$ to $\IMP(B)$ for $\CloneD_2\subseteq[B]$ by modifying 
    the reduction given in the proof of Lemma \ref{lem:implication-S00-and-S01}.
    
    Given a formula $\varphi$ in disjunctive normal form, 
    we define the formulae $\psi_1$ and $\psi_2$ as above.
    As $\CloneD_{2} \subseteq [B]$, we know that $g(x,y,z):=(x \land y) \lor (y \land z) \lor (x \land z) \in [B]$. Clearly, $g(x,y,\false)\equiv x \land y$ and $g(x,y,1)\equiv x \lor y$.
    Denote by $\psi_i^B(t,f)$, $i\in \{1,2\}$, the formula $\psi_i$ with
    all occurrences of $x \land y$ and $x \lor y$ replaced by a
    $B$-representation of $g(x,y,f)$ and $g(x,y,t)$, respectively,
    where $t$ and $f$ are new propositional variables. 
    Then $\psi_i^B(1,0)\equiv \psi_i$ and $\psi_i^B(0,1)\equiv \dual{\psi_i}$.
    The variables $x$ and $y$ occur several times in $g$, hence $\psi_1^B(t,f)$ and $\psi_2^B(t,f)$ might be
    exponential in the length of $\varphi$ (recall that $\psi_2$ is $\varphi$ with all negative literals 
    replaced by new variables). That this is not the case follows from the associativity of $\land$ and $\lor$: 
    we insert parentheses in such a way that $\psi_i^B$ can be transformed into a tree of logarithmic depth.
    
    We now map a pair $(\psi_1,\psi_2)$ to
    $(\psi_1',\psi_2')$ where 
    $$
      \psi_1' := g(\psi_1^B(t,f),t,f) \text{ and } \psi_2' := g(g(\psi_1^B(t,f),\psi_2^B(t,f),f),t,f).
    $$
    We claim that $\psi_1 \models \psi_2 \iff \psi_1' \models \psi_2'$.
    To verify this claim, let $\sigma$ be an arbitrary assignment for
    the variables in $\varphi$. Then $\sigma$ may be extended to
    $\{t,f\}$ in the following ways: 
    \begin{description}
      \item[$\sigma(t)=\true$ and $\sigma(f)=\false$:] This is the
        intended interpretation.  In this case,
        $g(\psi_1^B(\true,\false),\true,\false) \equiv \psi_1 \land \true \equiv
        \psi_1$ and $g(g(\psi_1^B(\true,\false),\psi_2^B(\true,\false),\false),\true,\false)
        \equiv (\psi_1 \land \psi_2) \land \true \equiv \psi_1
        \land \psi_2$. 
        Then $\psi_1 \models \psi_2$ iff $\psi_1 \models \psi_1 \land
        \psi_2$.

      \item[$\sigma(t)=\false$ and $\sigma(f)=\true$:] 
      In this case, we obtain that $g(\psi_1^B(\false,\true),\false,\true) \equiv \dual{\psi_1} \lor
      \false \equiv \dual{\psi_1}$ and
      $g(g(\psi_1^B(\false,\true),\psi_2^B(\false,\true),\true),\false,\true) \equiv (\dual{\psi_1}
      \lor \dual{\psi_2}) \lor \false \equiv \dual{\psi_1} \lor
      \dual{\psi_2}$. 
      As $\dual{\psi_1} \models \dual{\psi_1} \lor \dual{\psi_2}$
      is always valid, we conclude that $\psi_1'\models \psi_2'$ in this case.

      \item[$\sigma(t)=\sigma(f)=c$ with $c \in \{0,1\}$:] 
      Then both $\psi_1'$ and $\psi_2'$ are equivalent to $c$. Thus,
      as in the previous case, $\psi_1'\models\psi_2'$.
      
    \end{description}
    From this analysis, it follows that $\psi_1 \models
    \psi_2$ iff $\psi_1' \models \psi_2'$. 
    Hence, $\TAUTDNF \leqacm \IMP(B)$ via the reduction $\varphi \mapsto (\psi_1',\psi_2')$.
  \end{proof}

  \begin{lemma} \label{lemma_impl_L}
    Let $B$ be a finite set of Boolean functions such that $\CloneL_2 \subseteq [B]\subseteq \CloneL$.
    Then $\IMP(B)$ is $\ParityL$-complete under $\leqacm$-reductions.
  \end{lemma}
  
  \begin{proof} 
    Observe that $\Gamma \models \varphi$ iff  $\Gamma \cup \{\varphi \xor
    t,t\}$ is inconsistent, where $t$ is a new variable. 
    Let $\Gamma'$ denote $\Gamma \cup \{\varphi \xor t,t\}$ rewritten such that
    for all $\psi \in \Gamma'$, $\psi = c_0 \xor c_1x_1 \xor \cdots
    \xor c_nx_n$, where $c_0,\ldots,c_n \in \{\false,\true\}$. $\Gamma'$ is
    logspace constructible, since $c_0=\true$ iff $\psi(\false,\dots,\false)=\true$,
    and for $1\leq i \leq n$, $c_i=\true$ iff
    \[
      \psi(\false,\dots,\false) \not\equiv
      \psi(\underbrace{\false,\ldots,\false}_{i-1},\true,\false,\ldots,\false).
    \] 
    $\Gamma'$ can now be transformed into a system of linear equations $S$ via 
    \[
      c_0 \xor c_1x_1 \xor \cdots \xor c_nx_n \mapsto c_0 + c_1x_1 + \cdots + c_nx_n = 1 \pmod 2.
    \]
    Clearly, the resulting system of linear equations has a solution
    iff $\Gamma'$ is consistent. 
    The equations are furthermore defined over the field $\Z_2$, 
    hence existence of a solution can be decided in $\ParityL$~\cite{budaheme92}.

    For the $\ParityL$-hardness, note that solving a system of linear equations 
    over $\Z_2$ is indeed $\ParityL$-complete under $\leqacm$-reductions:
    let $\MODGAP_2$ denote the $\ParityL$-complete problem to decide
    whether a given directed acyclic graph $G$ with nodes $s$ and $t$
    has an odd number of distinct paths leading from $s$ to
    $t$. Buntrock et~al.\ \cite{budaheme92} give an $\NC{1}$-reduction from $\MODGAP_2$
    to the problem whether a given matrix over $\Z_2$ is non-singular.
    The given reduction is actually an $\AC{0}$ many-one reduction. 
    We reduce the latter problem to the complement of $\IMP(\{x\xor y \xor z\})$ and 
    then generalize the result to arbitrary finite sets $B$ such that $[B]=\CloneL_2$.
    The lower bound then follows from $\ParityL$ being closed under complement.

    First map the system $S$ of linear equations into a set of linear formulae $\Gamma$ via
    \[
     c_1x_1 + \cdots + c_nx_n = c \!\!\! \pmod 2 \ \mapsto \ c' \xor c_1x_1 \xor \cdots \xor c_nx_n,
   \]
    where $c' = \true$ if $c=0$, and $c' = \false$ otherwise. 
    Next replace the constant $\true$ with a fresh variable $t$, 
    pad all formulae having an even number of non-fictive variables with another fresh variable $f$,
    and let $\Gamma' := \Gamma \cup \{t\}$.
    We claim that $S$ has a solution iff $\Gamma' \not\models f$. 

    Suppose that $S$ has no solutions.
    If $\Gamma'$ is inconsistent, then $\Gamma' \models f$.
    Otherwise, $\Gamma'$ has a satisfying assignment $\sigma$. 
    Clearly, $\sigma(t)=\true$. 
    If $\sigma(f)=\false$, then $\Gamma'[t/\true,f/\false]$ is equivalent to $\Gamma$;
    hence the transformation of $\Gamma'[t/\true,f/\false]$ yields a system of linear equations $S'$ 
    that is equivalent to $S$ and that has a solution corresponding to $\sigma$\,---\,a contradiction 
    to our assumption. Thus $\sigma(f)=\true$ and, consequently, $\Gamma' \models f$.
    
    On the other hand, if $S$ has a solution, 
    then $\Gamma$ possesses a satisfying assignment $\sigma$
    with $\sigma(t)=\true$ and $\sigma(f)=\false$. 
    Again $\sigma\models \Gamma'$  iff $\sigma \models \Gamma$. 
    Hence, $\Gamma' \not\models f$.

    It remains to show that $x \xor y \xor z$ can be
    efficiently expressed in any set $B$ such that $[B]=\CloneL_2$,
    that is, 
    there exists a function $f_\xor \in [B]$ such that $f_\xor$ is 
    equivalent to $x \xor y \xor z$ and each variable occurs only once 
    in the body of $f_\xor$.
    Let $B$ be such that $[B]=\CloneL_2$ and let $g(x,y,z)$ be 
    a function from $[B]$ depending on three variables. Such a
    function $g$ exists because $x\oplus y \oplus z \in [B]=\CloneL_2$.
    As $g$ is a linear function, 
    replacing two occurrences of any variable with a fresh variable $t$ does not change $g$ modulo logical equivalence. 
    Let $n$ denote the number of occurrences of $x$ in $g$ and assume that $n$ is even. 
    Replacing all occurrences of $x$ with an arbitrary symbol yields a formula
    $g'(y,z)\equiv y\xor z \notin \CloneL_2$ which gives a contradiction.
    Analogous arguments hold for the number of occurrences of $y$ and $z$. 
    Hence, each of the variables $x$, $y$, and $z$ occurs an odd number of times, and
    replacing all but one occurrence of each $x$, $y$, and $z$ with $t$ yields a function $g'(x,y,z,t) \equiv x \xor y \xor z$ 
    in which each variable occurs exactly once.
  \end{proof}

  \begin{lemma} \label{lemma_impl_N}
    Let $B$ be a finite set of Boolean functions such that 
    $\CloneN_2 \subseteq [B]\subseteq \CloneN$.
    Then $\IMP(B)$ is contained in $\AC{0}\complexityClassFont{[2]}$ and $\MOD_2 \leqacm \IMP(B)$. 
  \end{lemma}
  
  \begin{proof}
    Let $B$ be a finite set of Boolean functions such that $\CloneN_2 \subseteq [B]\subseteq \CloneN$. 
    Let $\varphi$ be a $B$-formula and $\Gamma$ be a set of $B$-formulae, both over the set of propositions $\{x_1,\ldots,x_n\}$. 
    
    We will argue on membership in $\AC{0}\complexityClassFont{[2]}$ first.
    For all $f \in [B]$, $f$ is equivalent to some literal or a
    constant. Let $L := \{ l_i
    \mid \text{there exists } \psi \in \Gamma\colon l_i \equiv \psi \}$, 
    where $l_i = x_i$ or $l_i = \neg x_i$ for $1 \leq i \leq n$. 
    $L$ is computable from $\Gamma$ using an $\AC{0}$-circuit with oracle gates for $\MOD_2$: 
    for each formula in $\Gamma$, we determine the atom and count the number of preceding negations modulo 2.    
    In the case that $\Gamma$ is unsatisfiable, either $L=\emptyset$ or 
    there exist $l_i,l_j \in \Gamma$ with $l_i \equiv \lnot l_j$. 
    Both conditions can be checked in $\AC{0}$, 
    hence we may w.\,l.\,o.\,g.\ assume that $\Gamma$ is satisfiable.
    It now holds that
    \[
      \Gamma \models \varphi \iff \bigwedge_{l_i \in L} l_i \models
      \varphi \iff \mbox{for some } L' \subseteq L\colon \varphi \equiv \bigwedge_{l_i \in L'} l_i.
    \]
    It remains to compute an equivalent formula of the form
    $\bigwedge_{l_i \in L'} l_i$ from $\varphi$ and test whether $L'
    \subseteq L$. It is easy to see that the former task can again be
    performed in $\AC{0}\complexityClassFont{[2]}$, and the latter
    merely requires $\AC{0}$.  Thus we conclude $\IMP(B) \in
    \AC{0}\complexityClassFont{[2]}$. 
  
    For $\MOD_2 \leqacm \IMP(B)$, we claim that, for $w=w_1\cdots w_n\in \{0,1\}^n$, $w \in \MOD_2$ iff $t \models \neg^{w_1} \neg^{w_2} \cdots \neg^{w_n}  (\neg t)$, where $\neg^{\true} := \neg$, $\neg^{\false}:= \id$ and $t$ is a variable. 
    
    First observe that $t \models \neg^{w_1} \neg^{w_2} \cdots
    \neg^{w_n}  (\neg t)$ iff for all assignments $\sigma$ of $t$ to
    $\{\false,\true\}$, $\sigma \models t$ implies $\sigma
    \models  \neg^{w_1} \neg^{w_2} \cdots \neg^{w_n}  (\neg t)$. 
    Now, if $\sigma(t):=\false$, then $t \models \neg^{w_1} \neg^{w_2}
    \cdots \neg^{w_n}  (\neg t)$ is always true, whereas, if
    $\sigma(t):=\true$, then $t \models \neg^{w_1} \neg^{w_2} \cdots
    \neg^{w_n}  (\neg t)$ iff $\true \models \neg^{w_1} \neg^{w_2}
    \cdots \neg^{w_n} \false$. 
    Hence, the claim applies and $\MOD_2 \leqacm \IMP(B)$ follows.
  \end{proof}
  
  As an immediate consequence of the above lemma, we obtain the following corollary.
  
  \begin{corollary}
    Let $B$ be a finite set of Boolean functions such that 
    $\CloneN_2 \subseteq [B]\subseteq \CloneN$. 
    Then $\IMP(B)$ is $\AC{0}\complexityClassFont{[2]}$-complete under
    $\leqcd$-reductions.
  \end{corollary}
  
  \begin{lemma} \label{lem:implication-EV}
    Let $B$ be a finite set of Boolean functions such that $[B]
    \subseteq \CloneV$ or $[B] \subseteq \CloneE$.
    Then $\IMP(B)$ is in $\AC{0}$. 
  \end{lemma}
  
  \begin{proof}
    We prove the claim for $[B] \subseteq \CloneV$ only. The case $[B]
    \subseteq \CloneE$ follows analogously.
    
    Let $B$ be a finite set of Boolean functions such that $[B] \subseteq \CloneV$. 
    Let further $\Gamma$ be a finite set of $B$-formulae and let $\varphi$ be a
    $B$-formula such that $\Gamma$ and $\varphi$ only use the variables
    $x_1,\ldots,x_n$.
    Let $\varphi \equiv c_0 \lor c_1 x_1 \lor \cdots \lor c_n x_n$
    with constants $c_i \in \{\false,\true\}$ for $0 \leq i \leq n$.
    Equally, every formula from $\Gamma$ is equivalent to
    an expression of the form
    $c'_0 \lor c'_1 x_1 \lor \cdots \lor c'_n x_n$ with $c'_i \in \{\false,\true\}$.
    Then, $\Gamma \models \varphi$ iff either $c_0 = \true$ or there exists
    a formula $\psi\equiv c''_0 \lor c''_1 x_1 \lor \cdots \lor c''_n x_n$ from
    $\Gamma$ such that $c''_i \leq c_i$ for all $0 \leq i \leq n$ and $c''_i\in\{0,1\}$. 

    The value of $c_0$ can be determined by evaluating $\varphi(\false,\ldots,\false)$.
    Furthermore, for $1 \leq i \leq n$, $c_i=\false$ iff $c_0=\false$ and 
    \[
      \varphi(\underbrace{\false,\ldots,\false}_{i-1},\true,\false,\ldots,\false)=\false.
    \] 
    The values of the coefficients of formulae in $\Gamma$ can be computed analogously.
    Thus $\IMP(B)$ can be computed in constant depth using oracle
    gates for $B$-formula evaluation. 
    As $B$-formula evaluation is in $\NLOGTIME$ \cite{sch05} and
    $\NLOGTIME \subseteq \AC{0}$, the claim follows. 
  \end{proof}

\section{The Complexity of the Singleton-Premise Implication Problem}
  
  For a finite set $B$ of Boolean functions, we define
  the \emph{singleton-premise implication problem} for $B$-formulae as  
  
  \begin{center}
    \begin{tabular}{ll}
      \textit{Problem}:& $\IMP'(B)$ \\
      \textit{Instance}:  & Two $B$-formulae $\varphi$ and $\psi$.\\
      \textit{Question}: & Does $\varphi \models \psi$ hold?
    \end{tabular}
  \end{center}
  We classify the complexity of this problem as follows:
  \begin{theorem} \label{thm:imp=imp'}
    Let $B$ be a finite set of Boolean functions.
    Then $\IMP'(B) \in \AC{0}[2]$ and $\MOD_2 \leqacm \IMP'(B)$ if
    $\CloneL_2 \subseteq [B] \subseteq \CloneL$.
    For all other sets $B$, the problems $\IMP(B)$ and  
    $\IMP'(B)$ are equivalent. 
  \end{theorem}

  Before we prove Theorem~\ref{thm:imp=imp'}, let us try to give an intuitive explanation for
  the difference in the complexity of $\IMP'(B)$ for $\CloneL_2 \subseteq [B] \subseteq 
  \CloneL$ stems from.
  Deciding $\IMP(B)$ is equivalent to  solving a set of linear equations corresponding to the 
  set of premises. For $\IMP'(B)$, the premise is a single formula. 
  It hences suffice to determine whether there exists an assignment satisfying 
  the premise and setting to true an even (resp.\ odd) number of variables from the conclusion.
  
  \begin{proof}
    For $\CloneS_{00} \subseteq [B]$, $\CloneS_{10} \subseteq [B]$, and $\CloneD_2 \subseteq [B]$,
    observe that the proofs of Lemma \ref{lem:implication-S00-and-S01} and Lemma \ref{lem:implication-D2} actually establish $\co\NP$-hardness of $\IMP'(B)$.
    Analogously, for $\CloneN_2 \subseteq [B]$, $\MOD_2 \leqacm
    \IMP'(B)$ follows by the same reduction given in the proof of
    Lemma \ref{lemma_impl_N}. 
    For $[B] \subseteq \CloneV$ and $[B] \subseteq \CloneE$, we have $\IMP'(B) \leqacm \IMP(B) \in \AC{0}$. 
    It thus remains to show that $\IMP'(B) \in \AC{0}[2]$ for $[B] \subseteq \CloneL$, and 
    that $\MOD_2 \leqacm \IMP'(B)$ for $\CloneL_2 \subseteq [B]$.
    
    Let $(\varphi,\psi)$ be a pair of $B$-formulae over the variables $\{x_1,\ldots,x_n\}$.
    As $[B] \subseteq \CloneL$, $\varphi$ and $\psi$ are equivalent to expressions of the form
    $\varphi \equiv c_0 \xor c_1 x_1 \xor \cdots \xor c_n x_n$ and
    $\psi \equiv c_0' \xor c_1' x_1 \xor \cdots \xor c_n' x_n$, where
    $c_i,c_i' \in \{\false,\true\}$ for $1 \leq i \leq n$. 
    If $c_0=\dots=c_n=0$, then $\varphi\models\psi$ apparently
    holds.
    Therefore, let us assume that not all coefficients $c_i$ are 0.
    In this situation, we claim that $\varphi\models\psi$ is in fact
    equivalent to $\varphi \equiv \psi$.
    To prove this claim observe that
    $\varphi \models \psi$ iff 
    $$
     \chi := (c_0 \xor c_1 x_1 \xor \cdots
     \xor c_n x_n) \land (1 \xor  c_0' \xor c_1' x_1 \xor \cdots \xor
     c_n' x_n)
    $$ 
    is unsatisfiable. Let us assume now $\varphi \not\equiv \psi$.
    We will construct a satisfying assignment $\sigma$ for $\chi$.
    Let $I:=\{ i \in\{1,\dots, n\} \mid c_i=c_i' \}$ and
    define $\sigma(x_i):=\false$ for $i \in I$. 
    As $\varphi \not\equiv \psi$, the set $\overline{I} :=
    \{1,\ldots,n\}\setminus I$ is nonempty and for all $i \in
    \overline{I}$, $c_i=\true \iff c_i'=\false$. Hence, there is a
    partition $P_1 \uplus P_2 =\overline{I}$ such that 
    \[
      \sigma \models \chi \iff \sigma \models (c_0 \xor \bigoplus_{i \in P_1} c_i x_i) \land (1 \xor c_0' \xor \bigoplus_{i \in P_2} c_i' x_i).
    \]
    Here the subformulae $c_0 \xor \bigoplus_{i \in P_1} c_i x_i$ and
    $1 \xor c_0' \xor \bigoplus_{i \in P_2} c_i' x_i$ are over
    disjoint sets of variables. But still, both subformulae are
    satisfiable using an appropriate completion of
    $\sigma$. Consequently, $\sigma$ will also satisfy $\chi$ and hence
    the claim holds.
    
    Thus $\varphi \models \psi$ if either $c_0=\dots =c_n=0$ or
    $\varphi \equiv \psi$. 
    Similarly to the  proof of Lemma~\ref{lem:implication-EV}, it
    follows that the latter alternative holds 
    iff $c_i=c_i'$ for all $0 \leq i \leq
    n$. The coefficients $c_i$ can be determined from $c_0 = \varphi(\false,\ldots,\false)$ and 
    \[
      c_i=\varphi(\underbrace{\false,\ldots,\false}_{i-1},\true,\false,\ldots,\false)
      \xor c_0 
    \]
    for $ 1 \leq i \leq n$. 
    The values of the $c_i'$'s can be computed analogously. As $B$-formula evaluation is equivalent to $\MOD_2$ \cite{sch05} in this case, $\IMP'(B) \in \AC{0}[2]$. 
    
    It remains to prove $\MOD_2 \leqacm \IMP'(B)$ for $\CloneL_2
    \subseteq [B]$.
    Consider the mapping $h:\{0,1\}^\star \to \allFormulae(B)$,
    recursively defined by 
    \[
      h(x) = \begin{cases}
        f     & x=\varepsilon\\
        h(y)  & x=0y \\
        t \xor f \xor h(y) & x=1y
      \end{cases}
    \]
    where $\varepsilon$ denotes the empty word and $t,f$ are
    propositional variables.
    We claim that $x\mapsto (t,h(x))$ computes an $\leqacm$-reduction from $\MOD_2$ to
    $\IMP'(B)$. 
    To verify this claim, let $x \in \{0,1\}^\star$ be an instance of $\MOD_2$. 
    Then 
    \[
    \begin{array}{l@{\,}l@{\,}l}
      x \in \MOD_2    & \implies h(x) \equiv t & \implies (t,h(x)) \in \IMP'(B), \\
      x \notin \MOD_2 & \implies h(x) \equiv f & \implies (t,h(x)) \notin \IMP'(B).
    \end{array}
    \]
    Whence, $\MOD_2 \leqacm \IMP'(B)$ for $\CloneL_2 \subseteq [B]$.
  \end{proof}
  
  Let $\EQ(B)$ denote the equivalence problem for
  $B$-formulae. Obviously, $(\varphi,\psi) \in \EQ(B)$ iff
  $(\varphi,\psi) \in \IMP'(B)$ and $(\psi,\varphi) \in \IMP'(B)$.  
  As $\AC{0}$, $\AC{0}\complexityClassFont{[2]}$, and $\co\NP$ are all
  closed under intersection, we obtain as an immediate corollary a
  finer classification of the complexity of $\EQ$ than the one given by Reith \cite{rei03}. He
  establishes a dichotomy between $\co\NP$-hardness and membership in
  $\L$. We split the second case into two complexity degrees. 
  
  \begin{corollary} 
    Let $B$ be a finite set of Boolean functions.
    Then $\EQ(B)$ is $\co\NP$-complete under $\leqcd$-reductions 
    if $\CloneS_{00} \subseteq [B]$ or $\CloneS_{10} \subseteq [B]$ or 
    $\CloneD_2 \subseteq [B]$;  
    $\AC{0}\complexityClassFont{[2]}$-complete under
    $\leqcd$-reductions if $\CloneN_2 \subseteq [B] \subseteq \CloneN$; 
    and in $\AC{0}$ for all other cases.
  \end{corollary}

\section{Conclusion}

  In this paper we provided a complete classification of the
  complexity of the implication problem, $\IMP(B)$, and the singleton-premise
  implication problem, $\IMP'(B)$---fundamental problems in the area of
  propositional logic. 
  Though $\IMP'(B)$ is a restricted version of $\IMP(B)$,
  the simplification amounts to a difference for 
  $\CloneL_2 \subseteq [B] \subseteq \CloneL$ only: 
  $\IMP'(B)$ is $\AC{0}\complexityClassFont{[2]}$-complete under constant-depth reductions, 
  whereas $\IMP(B)$ is $\ParityL$-complete under $\AC{0}$ many-one reductions and thus strictly harder. 
  For all other clones, both problems have the same complexity.

  Due to the close relationship between the implication and the equivalence problem,
  we were also able to slightly refine the classification of the complexity of the 
  equivalence problem given in~\cite{rei03}.

\section*{Acknowledgements}

We thank the anonymous referees for
helpful comments and detailed suggestions on how to improve this
paper.

\bibliographystyle{plain}
\bibliography{thi-hannover}

\end{document}